\documentclass[pra, two column, show pacs]{revtex4}%
\usepackage{amsfonts}
\usepackage{amsmath}
\usepackage{amssymb}
\usepackage{graphicx}%
\providecommand{\U}[1]{\protect\rule{.1in}{.1in}}
\begin{document}
\title{Goos-H\"{a}nchen shifts in spin-orbit-coupled cold atoms}
\author{Lu Zhou$^{1}$\footnote{lzhou@phy.ecnu.edu.cn}, Jie-Li Qin$^{1}$, Zhihao
Lan$^{2,3}$\footnote{zhihao.lan@nottingham.ac.uk}, Guangjiong Dong$^{1}$, and Weiping
Zhang$^{1}$}
\affiliation{$^{1}$Department of Physics, State Key Laboratory of Precision Spectroscopy,
Quantum Institute for Light and Atoms, East China Normal University, Shanghai
200062, China}
\affiliation{$^{2}$Mathematical Sciences, University of Southampton, Highfield, Southampton}
\affiliation{$^{3}$Present address: School of Physics and Astronomy, University of Nottingham, Nottingham, NG7 2RD, UK}

\begin{abstract}
We consider a matter wave packet of cold atom gas impinging upon a step
potential created by the optical light field. In the presence of spin-orbit
(SO) coupling, the atomic eigenstates contain two types of evanescent states,
one of which is the ordinary evanescent state with pure imaginary wave vector
while the other possesses complex wave vector and is recognized as oscillating
evanescent state. We show that the presence and interplay of these two types
of evanescent states can give rise to two different mechanisms for total
internal reflection (TIR), and thus lead to unusual Goos-H\"{a}nchen (GH)
effect. As a result, not only large positive but also large negative GH shift
can be observed in the reflected atomic beam. The dependence of the GH shift
on the incident angle, energy and height of the step potential is studied numerically.

\end{abstract}

\pacs{03.75.-b, 42.25.Bs, 42.50.Gy}
\maketitle

Goos-Hanchen (GH) effect was first discovered in 1947 \cite{goosAP1949} as an
optical phenomenon in which a light beam incident upon the interface of two
media can experience a lateral shift under the condition of total internal
reflection (TIR). The origin of this lateral shift roots in the fact that the
incident beam is composed of different plane wave components with finite
transverse distribution, which will experience different phase shifts during
TIR. So the key physics inherent in GH effect is wave interference, and in
this sense it can also be generated with matter wave. As a matter of fact, GH
shift in graphene \cite{beenakkerPRL2009}, neutron \cite{haanPRL2010} and even
atom optics \cite{huangPRA2008} had been predicted.

Among these, electronic analogue of the GH effect had been carefully studied
\cite{chenJOA2013}. Compared to electromagnetic wave, electron provides extra
new degree of freedom for tuning the GH effect. For example, the effective
electron mass can be tuned from positive to negative \cite{dragomanJAP2007},
therefore the negative GH shift can be observed. In graphene
\cite{beenakkerPRL2009,chenEPJB2011}, for massless electrons in the
ultrarelativistic limit, it was found that the GH effect becomes dependent on
the spin degree of freedom. However, the electronic GH shift is about the
order of the electron de Broglie wavelength (on the scale of nanometer), which
impedes its direct observation in experiments. Recent years have witnessed
rapid advancement in laser cooling and trapping technology, which made
ultracold atoms with a relatively long de Broglie's wavelength available. As
such, phenomena originally predicted for electrons in the ultrarelativistic
limit, like zitterbewegung \cite{ZB} and Klein tunnelling \cite{Klein}, have
been successfully observed in experiments with cold atoms
\cite{Klein_Weitz,ZB_Spielman}, thus making cold atom systems appealing for
the study of the GH and relativistic effects \cite{relativistic_atom}.

Furthermore, recent experimental progress had also implemented spin-orbit (SO)
coupling in ultracold atomic gas by dressing cold atoms with lasers of special
configuration \cite{dalibardRMP2011}. In electronic systems, SO coupling can
originate from the interaction between the intrinsic electronic spin and the
magnetic field induced by its movement. It connects the electronic spin to its
orbital motion and thus the electron transport becomes spin-dependent. In the
presence of SO coupling, apart from the normal evanescent states, there exist
oscillating evanescent states \cite{sablikovPRB2007}, in which the matter wave
propagates with complex longitudinal wave vector. The role of the SO
interaction and evanescent states on the tunnelling dynamics of electron had
been studied previously \cite{sablikovPRB2007,banPRA2012}. However, the role
of these evanescent states on the TIR and thus GH effect has not been explored
to our knowledge. As we will show in the following, the presence and interplay
of these two types of evanescent states can give rise to two different
mechanisms for total internal reflection, and thus lead to unusual GH effect.

The Rashba SO coupling can be generated in neutral cold atoms with a tripod
scheme \cite{zhangPRL2012,JuzeliunasPRA2008,ZhangPRA2010}, in which the atomic
electronic energy level structure is composed of an excited state $\left\vert
0\right\rangle $ and three degenerate hyperfine ground-states $\left\vert
1\right\rangle $, $\left\vert 2\right\rangle $ and $\left\vert 3\right\rangle
$. The ground-state $\left\vert j\right\rangle $ $\left(  j=1,2,3\right)  $ is
coupled to $\left\vert 0\right\rangle $ via a laser field with Rabi frequency
$\Omega_{j}$. By appropriately designing the laser configurations
\cite{JuzeliunasPRA2008,ZhangPRA2010}, the atom-laser coupling system supports
two degenerate dark states, which are also the ground-states of the system. In
the subspace spaned by the dark states, the effective Hamiltonian can be
written as%
\begin{equation}
H=\frac{\hbar^{2}\mathbf{k}^{2}}{2m}+\hbar\alpha\left(  k_{x}\sigma_{y}%
-k_{y}\sigma_{x}\right)  +V\left(  x\right)  , \label{eq_hamiltonian}%
\end{equation}
with $\mathbf{k}^{2}=k_{x}^{2}+k_{y}^{2}$, $\alpha$ is the Rashba SO coupling
strength, $\sigma_x=[0,1;1,0]$, $\sigma_x=[0,-i;i,0]$ are the Pauli matrices acting on the two spin components and the scattering potential is described by $V\left(  x\right)
=V_{0}\Theta(x)$, where $\Theta(x)$ is the Heaviside step function. Such a
step potential can be created via a super-Gaussian laser beam with a
large-enough order \cite{SuperGaussian} and width compared to the atomic
de-Broglie wavelength.

In the case that $V\left(  x\right)  =V_0$ is a constant potential, the
eigenfunctions of Hamiltonian (\ref{eq_hamiltonian}) split into two branches
and can generally be expressed as%
\begin{equation}
\phi_{\mathbf{k}}^{\pm}=Ce^{i\left(  k_{x}x+k_{y}y\right)  }\binom{-2a\left(
k_{y}+ik_{x}\right)  }{E_{\mathbf{k}}^{\pm}-V_0-\mathbf{k}^{2}}, \label{eq_phi}%
\end{equation}
with $C$ the normalization constant and $a=m\alpha/\hbar$. $E_{\mathbf{k}%
}^{\pm}$ is the corresponding eigenenergy (scaled by $\hbar^{2}/2m$)
satisfying the relation
\begin{equation}
\left(  \mathbf{k}^{2}+V_0-E_{\mathbf{k}}\right)  ^{2}-4a^{2}\mathbf{k}^{2}=0,
\label{eq_energy}%
\end{equation}
from which we can get the energy spectrum $E_{\mathbf{k}}=k^{2}\pm2ak+V_0$ or
the modulus of the wave vector $k=\mp a+\sqrt{a^{2}+E_{\mathbf{k}}-V_0}$.

In the situation considered here, the system is left free along the
$y$-direction and semi-infinite in the $x$-direction, thus $k_{y}$ is real and
$k_{x}$ is generally complex. As those had been illustrated in
\cite{sablikovPRB2007}, the eigenfunctions of the system can be grouped into
three categories according to their properties: propagating states with
$k_{x}$ real, evanescent states (only exist near the boundary of the system
and propagate along it) with $k_{x}=i\kappa$ (requiring $\left\vert
\kappa\right\vert <\left\vert k_{y}\right\vert $) and oscillating evanescent
states with $k_{x}=K_{x}^{\prime}+iK_{x}^{\prime\prime}$, in which
$K_{x}^{\prime}$, $K_{x}^{\prime\prime}$ satisfy: $K_{x}^{\prime2}%
K_{x}^{\prime\prime2}=a^{2}\left(  V_0-E-a^{2}\right)  $ and $K_{x}^{\prime
2}-K_{x}^{\prime\prime2}=2a^{2}+E-V_0-k_{y}^{2}$. It is clear that in order for
the oscillating evanescent states to exist, the condition of $V_0>a^{2}+E$ needs
to be satisfied.

\begin{figure}[h]
\includegraphics[width=8cm]{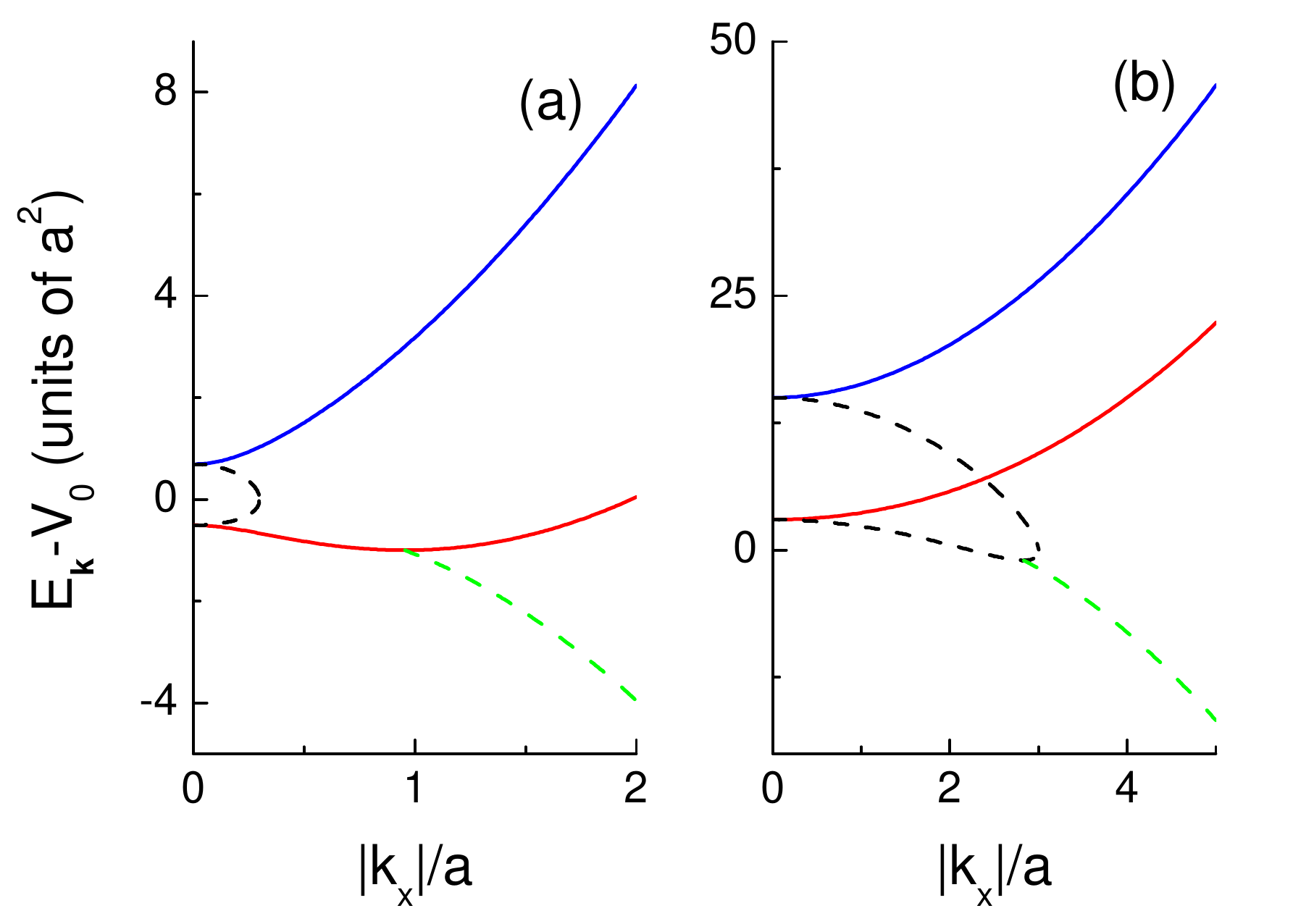}\caption{{\protect\footnotesize (Color
online) Energy spectra of the different states described by $E_{\mathbf{k}%
}=k^{2}\pm2ak+V_0$. The solid (blue and red) lines correspond to the up and down
branches of the propagating states. The black dashed line represents the
evanescent states while the green (light) dashed line is for the oscillating
evanescent states. The parameters are set as (a) }$k_{y}=0.3a$%
{\protect\footnotesize ; (b) }$k_{y}=3a${\protect\footnotesize .}}%
\label{fig_energy}%
\end{figure}

\begin{figure}[h]
\includegraphics[width=8cm]{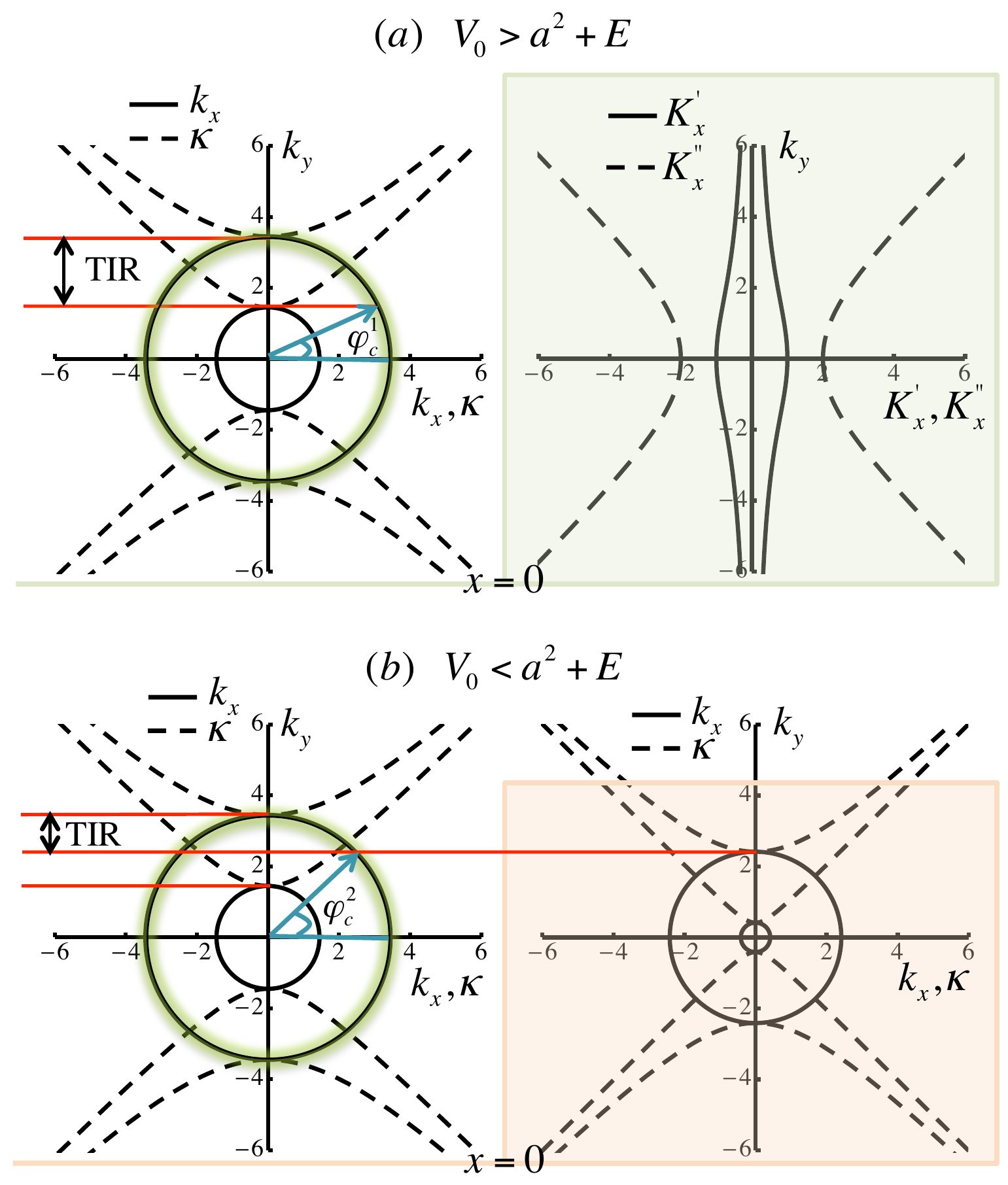}\caption{{\protect\footnotesize (Color online)
Schematic to illustrate the two different mechanisms for TIR when the incident
beam is prepared in the }$k=a+\sqrt{a^{2}+E}$ {\protect\footnotesize branch of
the propagating states (outer circles in the left). The different eigenstates
in (a) and (b) are plotted with }$E=5a^{2},V_{0}=10a^{2}$
{\protect\footnotesize and }$E=5a^{2},V_{0}=4a^{2}$
{\protect\footnotesize respectively, however, the figure is for illustrative
purposes only. In (a), where }$V_{0}>a^{2}+E$ {\protect\footnotesize such that
only oscillating evanescent states exist inside the step potential, the range
of }$k_{y}$ {\protect\footnotesize that gives TIR as marked by the arrows is
only determined by the incident energy }$E$ {\protect\footnotesize of the beam
only. In (b), where }$V_{0}<a^{2}+E$ {\protect\footnotesize and consequently
no oscillating evanescent state exists inside the step potential, the TIR
happens only when evanescent states are available at both sides of the
potential. As a result, the range of }$k_{y}$ {\protect\footnotesize that
gives TIR has an additional constraint from the potential compared with the
case in (a) and the critical angle for TIR to happen crucially depends on the
height of the potential. }}%
\label{TIR}%
\end{figure}

In order to better understand the properties of these eigenstates, we plot in
Fig. \ref{fig_energy} the energy spectra ($E_{\mathbf{k}}-V_0$) as a function of
$\left\vert k_{x}\right\vert $ for two typical values of $k_{y}$, which
exhibit different structures depending on the value of $k_{y}$. For both
cases, the two branches of propagating states are separated by a gap of
$4a\left\vert k_{y}\right\vert $ at $\left\vert k_{x}\right\vert =0$. When
$\left\vert k_{y}\right\vert <a$, $\left\vert k_{x}\right\vert =\sqrt
{a^{2}-k_{y}^{2}}$ is the energy minimum of the lower propagating branch and
the dispersion curve of the evanescent states forms a lobe with its tip
located at $\left\vert k_{x}\right\vert =\left\vert k_{y}\right\vert $, which
intersects with the energy spectra of propagating states at $\left\vert
k_{x}\right\vert =0$. While for $\left\vert k_{y}\right\vert >a$, $\left\vert
k_{x}\right\vert =0$ becomes the energy minimum of the lower propagating
branch, which intersects with the evanescent lobe at some finite $\left\vert
k_{x}\right\vert $ besides $\left\vert k_{x}\right\vert =0$. The oscillating
evanescent states possess minimum energies among these three types of
solutions for both cases, and they are linked to the energy minimum of the
lower propagating branch for $\left\vert k_{y}\right\vert <a$ and the
evanescent lobe for $\left\vert k_{y}\right\vert >a$.

The presence of two different types of evanescent states provides new
possibilities for TIR and could has a crucial effect on the resulting GH
shift. It would be convenient to discuss the conditions for TIR to happen
before we move on to the calculations and discussions of the GH shift. First,
the energy spectra in Fig. \ref{fig_energy} indicate that, in order for TIR to
take place, the incident atomic beam should be prepared in the lower
dispersion branch of propagating states or the ``outer circles" in the left of
Fig. \ref{TIR}. This is because any propagating states prepared in the upper
branch (or ``inner circle") will always lead to double reflection
\cite{juzeliunasPRL2008}, instead of TIR. Second, in the normal case without
SO coupling, there will be pure evanescent wave along the inner boundary of
the step potential when TIR takes place. The GH shift in this case can then be
understood as to account for the penetration of the evanescent field to the
other side of the interface and as such, the GH shift should be positive. The
new physics inherent in the present case with SO coupling can be understood as
there are two different mechanisms for TIR to take place.

(i) when $V_{0}>a^{2}+E$, the only available states inside the step potential
are oscillating evanescent ones. So the TIR happens when one of the reflected
beams at the $x<0$ side is evanescent, see the marked range of $k_{y}$ by
arrows in Fig. \ref{TIR}a, the lower bound of which defines the critical
incident angle $\varphi_{c}^{1}$ of the TIR
\begin{equation}
\varphi_{c}^{1}=\sin^{-1}\frac{\sqrt{E+a^{2}}-a}{\sqrt{E+a^{2}}+a},
\label{eq_c1}%
\end{equation}
which is independent of the height of the step potential. The available
oscillating evanescent states inside the step potential are composed of two
counter-propagating ones ($\pm K_{x}^{^{\prime}}$, see the right of Fig.
\ref{TIR}a), one of which penetrates along the potential boundary and gives
rise to evanescent wave on the outer boundary of the step potential. This
accounts for the creation of negative GH shifts.

(ii) when $V_{0}<a^{2}+E$ such that no oscillating evanescent state exists
inside the step potential, the TIR occurs only when evanescent states are
available at both sides of the potential. Thus apart from the marked range of
$k_{y}$ as in the case of $V_{0}>a^{2}+E$, there is an extra constraint from
the potential side. The lower bound of the overlapped range as marked in Fig.
\ref{TIR}b defines the critical incident angle $\varphi_{c}^{2}$ for the TIR
to happen in this case,
\begin{equation}
\varphi_{c}^{2}=\sin^{-1}\frac{\sqrt{E-V_{0}+a^{2}}+a}{\sqrt{E+a^{2}}+a},
\label{eq_c2}%
\end{equation}
and in this case the critical angle shows a crucial dependence on the height
of the step potential.

After the discussion of the two different mechanisms for TIR to occur, we now
proceed to the calculations and discussions of the resulting GH shifts. The
atoms are initially prepared in the $k=a+\sqrt{a^{2}+E}$ branch of the
propagating states with energy $E$ and incident upon the step potential from
$x<0$, the wavefunction reads%
\begin{equation}
\Psi_{in}=\int dk_{y}f\left(  k_{y}-k_{y0}\right)  e^{i\left(  k_{x}%
x+k_{y}y\right)  }\binom{-e^{-i\varphi\left(  k_{y}\right)  /2}}%
{ie^{i\varphi\left(  k_{y}\right)  /2}}, \label{eq_incident}%
\end{equation}
in which the distribution function $f\left(  k_{y}-k_{y0}\right)  $ describes
that the incident atomic beam has a narrow angular distribution centering at
$k_{y}=k_{y0}$ with incident angle $\varphi\left(  k_{y}\right)  =\arg\left(
k_{x}+ik_{y}\right)  $. We can then expand $\varphi\left(  k_{y}\right)  $
around $k_{y}=k_{y0}$ and as such $\varphi\left(  k_{y}\right)  \simeq
\varphi\left(  k_{y0}\right)  +\varphi^{\prime}\left(  k_{y0}\right)  k_{y}$.
By using the Fourier transform shift theorem, i.e., a linear phase shift in
the wave-vector domain introduces a translation in the space domain, the
centre of the incident atomic beam for the two spin components at $x=0$ will
locate at $y_{\pm}^{in}=\pm\varphi^{\prime}\left(  k_{y0}\right)  /2$ respectively.

Under the condition of TIR, from Fig.\ref{TIR} we know that the reflected wave
is a linear superposition of two waves: One is the reflected propagating wave%
\begin{equation}
\Psi_{r}=\int dk_{y}f\left(  k_{y}-k_{y0}\right)  e^{i\left(  -k_{x}%
x+k_{y}y\right)  }r\left(  k_{y}\right)  \binom{ie^{i\varphi\left(
k_{y}\right)  /2}}{-e^{-i\varphi\left(  k_{y}\right)  /2}}, \label{eq_reflect}%
\end{equation}
which is obtained from Eq. (\ref{eq_incident}) by replacing $k_{x}%
\rightarrow-k_{x}$, $\varphi\rightarrow\pi-\varphi$. And $r\left(
k_{y}\right)  =\left\vert r\left(  k_{y}\right)  \right\vert e^{i\phi\left(
k_{y}\right)  }$ is the reflection amplitude. The other wave on the incident
side is an evanescent one%
\begin{equation}
\Psi_{e}=\int dk_{y}f\left(  k_{y}-k_{y0}\right)  e^{\kappa x+ik_{y}y}s\left(
k_{y}\right)  \binom{-\frac{2a(\kappa+k_{y})}{E-k_{y}^{2}+\kappa^{2}}}{1},
\label{eq_evanscent}%
\end{equation}
in which $s$ is the reflection amplitude of the evanescent wave with
$\kappa>0$. In the case of TIR, $\left\vert r\left(  k_{y}\right)  \right\vert
=1$ becomes independent of $k_{y}$, then at the interface $x=0$ the centre of
the two reflected atomic components will be at $y_{\pm}^{r}=-\phi^{\prime
}\left(  k_{y0}\right)  \mp\varphi^{\prime}\left(  k_{y0}\right)  /2$.
Combining the expressions of $y_{\pm}^{r}$ and $y_{\pm}^{in}$, the atomic beam
will experience a lateral shift upon TIR: $\sigma_{\pm}=y_{\pm}^{r}-y_{\pm
}^{in}=-\phi^{\prime}\left(  k_{y0}\right)  \mp\varphi^{\prime}\left(
k_{y0}\right)  $ and the average shift
\begin{equation}
\sigma=\left(  \sigma_{+}+\sigma_{-}\right)  /2=-\phi^{\prime}\left(
k_{y0}\right)  \label{eq_gh}%
\end{equation}
is recognized as the GH shift.

As we have discussed above, under the condition of TIR the states in the step
potential can either be oscillating evanescent states ($V_{0}>a^{2}+E$) or
evanescent states ($V_{0}<a^{2}+E$). The theoretical derivations of the
reflection amplitudes in the two cases are similar, i.e., first write down the
wavefunction inside the step potential and then match the wavefunction using
the boundary conditions at $x=0$. In the following, we will focus on the
derivation when the oscillating evanescent states are the only available ones
\cite{note}, such that the waves in the step potential have the following form%
\begin{align}
\Psi_{t}  &  =\int dk_{y}f\left(  k_{y}-k_{y0}\right)  e^{-K_{x}^{\prime
\prime}x+ik_{y}y}\left[  b_{1}\left(  k_{y}\right)  \binom{a\frac
{iK_{x}^{\prime}-K_{x}^{\prime\prime}+k_{y}}{a^{2}+iK_{x}^{\prime}%
K_{x}^{\prime\prime}}}{1}\right. \nonumber\\
&  \left.  \times e^{iK_{x}^{\prime}x}+b_{2}\left(  k_{y}\right)
e^{-iK_{x}^{\prime}x}\binom{a\frac{-iK_{x}^{\prime}-K_{x}^{\prime\prime}%
+k_{y}}{a^{2}-iK_{x}^{\prime}K_{x}^{\prime\prime}}}{1}\right]  ,
\label{eq_oscillating}%
\end{align}
in which $K_{x}^{\prime}$ and $K_{x}^{\prime\prime}$ are positive real and
$b_{1\left(  2\right)  }$ are the transmission amplitudes.

By integrating the Schr\"{o}dinger equation $H\Psi(x)=E\Psi(x)$ over the
interval expanded around the interface $x=0$, we get%
\begin{align}
\Psi_{in}|_{x=0}+\Psi_{r}|_{x=0}+\Psi_{e}|_{x=0}  &  =\Psi_{t}|_{x=0}%
,\nonumber\\
\frac{\partial\Psi_{in}}{\partial x}|_{x=0}+\frac{\partial\Psi_{r}}{\partial
x}|_{x=0}+\frac{\partial\Psi_{e}}{\partial x}|_{x=0}  &  =\frac{\partial
\Psi_{t}}{\partial x}|_{x=0}, \label{eq_boundary}%
\end{align}
From Eqs. (\ref{eq_boundary}), the scattering index $\left\{  r,s,b_{1}%
,b_{2}\right\}  $ can be found to satisfy the following matrix equation%
\begin{equation}
\mathcal{M}\left(
\begin{array}
[c]{c}%
r\\
s\\
b_{1}\\
b_{2}%
\end{array}
\right)  =\left(
\begin{array}
[c]{c}%
e^{-i\varphi/2}\\
-ie^{i\varphi/2}\\
ik_{x}e^{-i\varphi/2}\\
k_{x}e^{i\varphi/2}%
\end{array}
\right)  , \label{eq_app_linear}%
\end{equation}
with \begin{widetext}%
\begin{equation}
\mathcal{M}=\left(
\begin{array}
[c]{cccc}%
ie^{i\varphi/2} & -2a\frac{\kappa+k_{y}}{E-k_{y}^{2}+\kappa^{2}} &
-a\frac{iK_{x}^{\prime}-K_{x}^{\prime\prime}+k_{y}}{a^{2}+iK_{x}^{\prime}%
K_{x}^{\prime\prime}} & -a\frac{-iK_{x}^{\prime}-K_{x}^{\prime\prime}+k_{y}%
}{a^{2}-iK_{x}^{\prime}K_{x}^{\prime\prime}}\\
-e^{-i\varphi/2} & 1 & -1 & -1\\
k_{x}e^{i\varphi/2} & -2a\kappa\frac{\kappa+k_{y}}{E-k_{y}^{2}+\kappa^{2}} &
-a\left(  iK_{x}^{\prime}-K_{x}^{\prime\prime}\right)  \frac{iK_{x}^{\prime
}-K_{x}^{\prime\prime}+k_{y}}{a^{2}+iK_{x}^{\prime}K_{x}^{\prime\prime}} &
a\left(  iK_{x}^{\prime}+K_{x}^{\prime\prime}\right)  \frac{-iK_{x}^{\prime
}-K_{x}^{\prime\prime}+k_{y}}{a^{2}-iK_{x}^{\prime}K_{x}^{\prime\prime}}\\
ik_{x}e^{-i\varphi/2} & \kappa & -iK_{x}^{\prime}+K_{x}^{\prime\prime} &
iK_{x}^{\prime}+K_{x}^{\prime\prime}%
\end{array}
\right)  ,\label{eq_app_matrix}%
\end{equation}
\end{widetext}
from which the reflection amplitude $r$ can be derived straightforwardly as%
\begin{equation}
r=\frac{e^{-i\varphi/2}A+ie^{i\varphi/2}B+ik_{x}e^{-i\varphi/2}C-k_{x}%
e^{i\varphi/2}D}{ie^{i\varphi/2}A+e^{-i\varphi/2}B+k_{x}e^{i\varphi/2}%
C-ik_{x}e^{-i\varphi/2}D}, \label{eq_app_reflect}%
\end{equation}
where $A,B,C,D$\ are minors of the first column entries of matrix
$\mathcal{M}$. \begin{figure}[h]
\includegraphics[width=8cm]{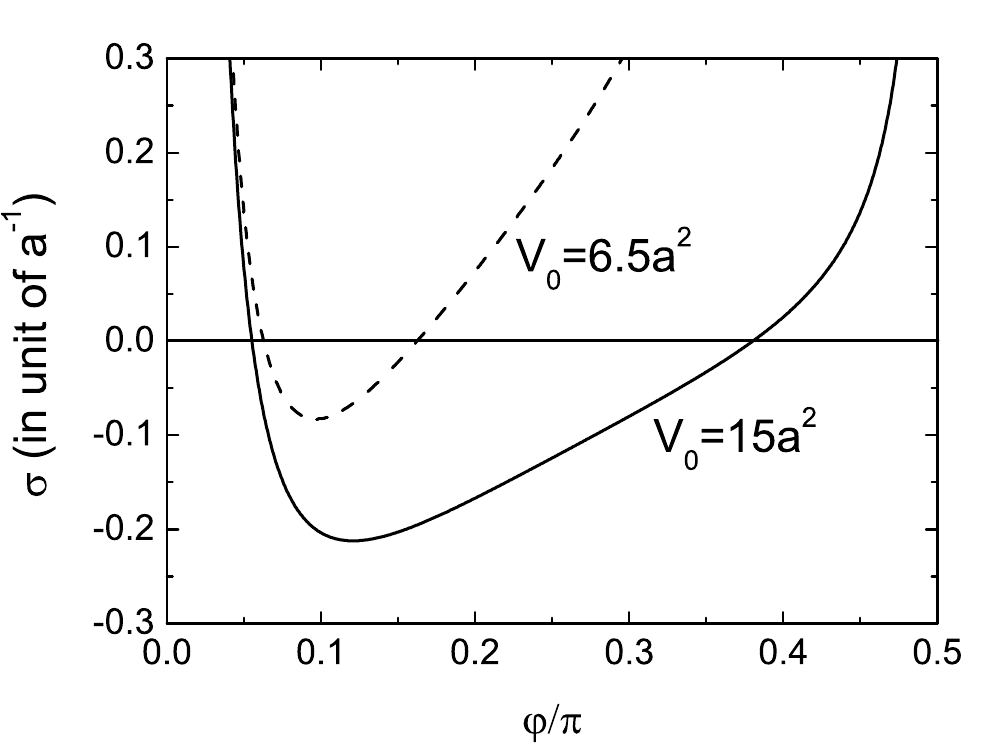}\caption{{\protect\footnotesize The GH shift
}$\sigma$ {\protect\footnotesize versus the incident angle }$\varphi$
{\protect\footnotesize for }$E=0.1a^{2}$ {\protect\footnotesize and }%
$V_{0}=15a^{2}${\protect\footnotesize (solid line), }$V_{0}=6.5a^{2}$
{\protect\footnotesize (dashed line).}}%
\label{fig_gh shift}%
\end{figure}

\begin{figure}[h]
\includegraphics[width=8cm]{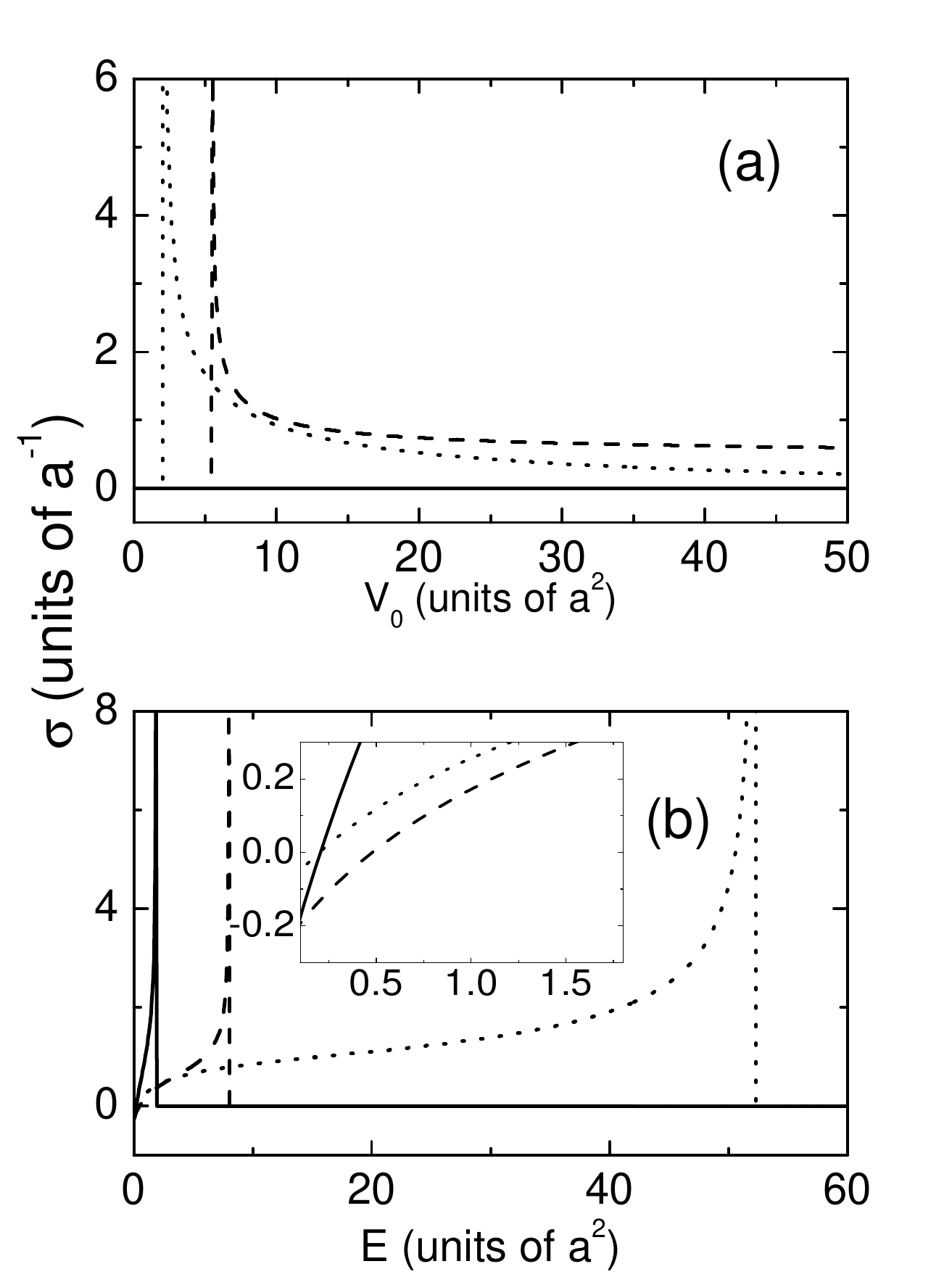}\caption{{\protect\footnotesize (a) The GH
shift }$\sigma$ {\protect\footnotesize versus the potential height }$V_{0}$
{\protect\footnotesize for }$E=5a^{2}$ {\protect\footnotesize and (b) }%
$\sigma${\protect\footnotesize versus the incident energy }$E$
{\protect\footnotesize for }$V_{0}=15a^{2}${\protect\footnotesize at incident
angle }$\varphi=\pi/12$ {\protect\footnotesize (solid line), }$\varphi=\pi/6$
{\protect\footnotesize (dashed line) and }$\varphi=\pi/3$
{\protect\footnotesize (dotted line). The inset figure display the occurance
of negative GH shift for small incident energy.}}%
\label{fig_energy and potential}%
\end{figure}

By calculating the phase shift inherent in the reflection amplitude
$r=e^{i\phi(k_{y})}$ according to Eq. (\ref{eq_gh}) under the condition of
TIR, GH shift can be derived and the results are shown in Fig.
\ref{fig_gh shift} and Fig. \ref{fig_energy and potential}. The GH shift is
scaled in unit of $a^{-1}$, which is the wavelength of the light field used to
create the SO coupling and is on the order of several hundred nanometres. This
is much larger than the electronic de Broglie wavelength and could be easily
observed in cold atom experiments. From Fig. \ref{fig_gh shift} one can see
that when the incident angle exceeds the critical angle ($0.0076\pi$), TIR
will take place. This critical angle is independent of the step potential
height $V_{0}$, resulting from the fact that $V_{0}>E+a^{2}$. From Fig.
\ref{fig_gh shift} one can also observe the appearance of negative GH shift.
With the increase of the incident energy $E$, the GH shift will gradually
become positive.

The dependence of the GH shifts on the height of the potential $V_{0}$ and the
incident energy $E$ are shown in Fig. \ref{fig_energy and potential}. Fig.
\ref{fig_energy and potential}(a) verifies that TIR occurs only when Eq.
(\ref{eq_c2}) is satisfied. For examples, when the incident energy is
$E=5a^{2}$, according to Eq. (\ref{eq_c2}), for incident angles of $\pi/6$ and
$\pi/3$, the critical heights of the step potential for the TIR to occur are
$5.47a^{2}$ and $2.05a^{2}$ respectively, which show good agreements with the
numerics in Fig. \ref{fig_energy and potential}(a). On the other hand, Fig.
\ref{fig_energy and potential}(b) indicates that negative GH shift can only be
observed for very small incident energy and the TIR will take place in a
larger energy interval with larger incident angle. Again, the theoretical
critical energies for TIR to happen according to Eq. (\ref{eq_c1}) and
(\ref{eq_c2}) with incident angles of $\pi/12$, $\pi/6$ and $\pi/3$ are
$1.88a^{2}$, $8.0a^{2}$ and $52.3a^{2}$ respectively and from Fig.
\ref{fig_energy and potential}(b) we see that they agree very well with the numerics.

In summary, we studied the GH effect in cold atoms with SO coupling and found
that there are two different mechanisms for TIR to occur, which can lead to
unusual GH shifts, e.g., not only large positive but also large negative GH
shifts can be observed. In addition, the modulation of the GH shift can be
realized by changing the potential height and the incident energy. Since the GH shift is much larger than the
atomic de Broglie wavelength, one can expect that it can  readily be measured in experiment from the density evolution of the atomic ensemble via
absorption imaging \cite{detect}.

This work is supported by the National Basic Research Program of China (973
Program) under Grant No. 2011CB921604, the National Natural Science Foundation
of China under Grant No. 11374003, 91436211, 11234003 and 11129402. Z.L.
acknowledges support from EPSRC grant No.EP/I018514/1.


\begin{thebibliography}{99}
\bibitem {note}In practice, it is not necessary to derive the reflection
amplitude when $V_0<a^{2}+E$ since the reflection amplitude in this case is
analytically connected to the one of $V_0>a^{2}+E$ in the complex plane.
\end{thebibliography}
\end{document}